\documentclass[sigconf]{acmart} 

\AtBeginDocument{%
  }

\copyrightyear{2025}
\acmYear{2025}
\setcopyright{cc}
\setcctype{by}
\acmConference[LiveRAG@SIGIR 2025]%
  {SIGIR 2025 LiveRAG Challenge, held in conjunction with the 48th International ACM SIGIR Conference on Research and Development in Information Retrieval (SIGIR 2025)}%
  {July 13--17, 2025}%
  {Padua, Italy}
\acmBooktitle{SIGIR 2025 LiveRAG Challenge, held in conjunction with the 48th International ACM SIGIR Conference on Research and Development in Information Retrieval (SIGIR 2025), July 13--17, 2025, Padua, Italy}
\acmDOI{}
\acmISBN{}

\begin{document}

\title{RAGentA: Multi-Agent Retrieval-Augmented Generation for Attributed Question Answering}

\author{Ines Besrour}
\authornote{Both authors contributed equally to this research.}
\email{ines.besrour@mailbox.tu-dresden.de}
\affiliation{%
  \institution{TU Dresden \& ScaDS.AI Dresden/Leipzig}
  \city{Dresden}
  \country{Germany}}
\orcid{0009-0007-2736-8971}

\author{Jingbo He}
\authornotemark[1]
\email{jingbo.he@mailbox.tu-dresden.de}
\affiliation{%
  \institution{TU Dresden \& ScaDS.AI Dresden/Leipzig}
  \city{Dresden}
  \country{Germany}}
\orcid{0009-0000-3653-2840}

\author{Tobias Schreieder}
\email{tobias.schreieder@tu-dresden.de}
\affiliation{%
  \institution{TU Dresden \& ScaDS.AI Dresden/Leipzig}
  \city{Dresden}
  \country{Germany}}
\orcid{0009-0000-8268-4204}

\author{Michael Färber}
\email{michael.faerber@tu-dresden.de}
\affiliation{%
  \institution{TU Dresden \& ScaDS.AI Dresden/Leipzig}
  \city{Dresden}
  \country{Germany}}
\orcid{0000-0001-5458-8645}

\renewcommand{\shortauthors}{Ines Besrour, Jingbo He, Tobias Schreieder, and Michael Färber}

\begin{abstract}
  We present RAGentA, a multi-agent retrieval-augmented generation (RAG) framework for attributed question answering (QA) with large language models (LLMs). With the goal of trustworthy answer generation, RAGentA focuses on optimizing answer correctness, defined by coverage and relevance to the question and faithfulness, which measures the extent to which answers are grounded in retrieved documents. RAGentA uses a multi-agent architecture that iteratively filters retrieved documents, generates attributed answers with in-line citations, and verifies completeness through dynamic refinement. Central to the framework is a hybrid retrieval strategy that combines sparse and dense methods, improving Recall@20 by 12.5\% compared to the best single retrieval model, resulting in more correct and well-supported answers. Evaluated on a synthetic QA dataset derived from the FineWeb index, RAGentA outperforms standard RAG baselines, achieving gains of 1.09\% in correctness and 10.72\% in faithfulness. These results demonstrate the effectiveness of our multi-agent RAG architecture and hybrid retrieval strategy in advancing trustworthy QA with LLMs.
\end{abstract}



\maketitle

\section{Introduction}
\label{introduction}

LLMs are increasingly applied to tasks such as multi-step reasoning~\cite{Wei2022CoT}, summarization~\cite{Zhang2024Summarization}, and open-domain QA~\cite{Singhal2025QA}. Nevertheless, LLMs often generate hallucinations, meaning factually incorrect or misleading outputs~\cite{Huang2025Hallucination}. Because users currently have limited means to verify LLM-generated answers, hallucinated information becomes especially problematic. This lack of verifiability poses a significant challenge for deploying LLMs in high-stakes domains such as science, healthcare, and law, where factuality and trust are essential.

RAG addresses this issue by combining traditional information retrieval with LLMs~\cite{Lewis2020RAG}. Relevant documents are retrieved from external sources and used to guide the model's generation. This approach reduces the likelihood of unsupported claims by grounding answers in curated, up-to-date information. Building on this foundation, \citet{Gao2023ALCE} show that RAG can also enable attribution, for example via in-line citations. This allows users to verify the sources of generated content and increases trustworthiness.

As part of our participation in the SIGIR 2025 LiveRAG Challenge, we propose a multi-agent RAG framework for attributed QA that emphasizes both answer correctness and faithfulness. We extend the MAIN-RAG framework~\cite{Chang2024MAIN-RAG}, which employs three agents to score the relevance of retrieved documents for improved answer correctness. We enhance this architecture by introducing a novel agent that improves the faithfulness of the output through fine-grained attribution and refines the LLM answer when needed. This collaborative setup helps close the gap toward more trustworthy LLM applications for QA. Our key contributions are as follows:
\begin{itemize}
\item We propose \textit{RAGentA}, a collaborative multi-agent RAG framework that improves answer faithfulness through fine-grained citations, especially for multi-source questions.
\item We ensure high retrieval quality through a hybrid sparse-dense approach, reinforced by agent-based relevance scoring to select the most suitable documents for generation.
\item We test RAGentA against a baseline RAG approach by building a diverse synthetic QA dataset from the FineWeb index to assess retrieval, answer correctness, and faithfulness.
\end{itemize}

\section{Related Work}
\label{related_work}

\textbf{Multi-Agent RAG.} Recent developments in RAG have increasingly leveraged multi-agent systems to improve performance and scalability in QA tasks. \citet{Zhao2024Longagent} introduced \textit{LongAgent}, a collaborative multi-agent framework designed to enable QA over very long documents. In this approach, queries are decomposed into sub-tasks handled by individual agents, each processing a segment of the input. A leader agent coordinates their outputs and synthesizes the final answer.
Building on the idea of agent specialization, \citet{Zhu2024ATM} proposed \textit{ATM}, a dual-agent adversarial RAG framework consisting of a generator and an attacker. The attacker injects challenging distractor documents, forcing the generator to distinguish relevant information even in misleading contexts through adversarial training.
Expanding the agent interaction paradigm, \citet{Yang2024IM-RAG} introduced \textit{IM-RAG}, which models an “inner monologue” among a reasoner, retriever, and refiner. These agents engage in iterative communication to refine both the queries and the retrieved content before generating a final answer.
Further emphasizing task specialization, \citet{Jang2024AU-RAG} developed \textit{AU-RAG}, where distinct agents are dedicated to sub-tasks such as query understanding, document retrieval, and answer generation. Similarly, \textit{AgentFusion} by \citet{Saeid2024AgentFusion} assigns specific roles such as planning, verification, and refinement to multiple agents within the generation pipeline, thereby enabling a more modular and robust generation process.
\citet{Chang2024MAIN-RAG} proposed \textit{MAIN-RAG}, a training-free, multi-agent architecture that focuses explicitly on document filtering and evidence scoring. This approach significantly improves answer accuracy while mitigating the impact of noisy retrievals.
Our proposed framework, \textit{RAGentA}, builds directly on the insights from MAIN-RAG, aiming to further enhance retrieval robustness and by providing attribution to sources used for text generation.

\textbf{Attributed Text Generation.} Also known as citation generation, this task aims to produce text with explicit citations to source documents, thereby enhancing trustworthiness. As described by \citet{Huang2024Citation}, LLM attribution is broadly categorized into parametric and non-parametric approaches. \textit{Parametric} approaches, such as Galactica~\citep{Taylor2022Galactica}, generate citations only using the model's internal knowledge. In contrast, \textit{non-parametric} methods leverage external knowledge sources during generation to provide attribution. Non-parametric approaches can be divided into post-generation and post-retrieval paradigms. \textit{Post-generation} approaches first generate an answer, then identify supporting evidence. For instance, RARR~\citep{Gao2023RARR} detects unsupported claims in generated answers and revises them using retrieved documents. \citet{Ramu2024PostHoc} improve attribution in long documents by decomposing answers into coarse-grained segments to better align with evidence. \citet{Huang2024CitationReward} propose fine-grained reward models for training LLMs to generate accurate citations. In this work, we focus on \textit{post-retrieval} approaches that follow the RAG paradigm, retrieving evidence prior to answer generation. \citet{Gao2023ALCE} introduced an initial approach and ALCE, a benchmark dataset widely used to evaluate attribution. \citet{Fierro2024LearningToPlan} propose a planning-based strategy, where models generate citation plans before answering. \citet{Berchansky2024CoTAR} develop CoTAR, applying chain-of-thought reasoning with hierarchical citation to improve attribution granularity. \citet{Qi2024ModelInternals} present MIRAGE, which uses internal model representations to trace answer spans back to supporting passages, improving faithfulness. Finally, \citet{Patel2024MultiSource} demonstrate a method for generating multiple citations per sentence, enabling more nuanced and comprehensive attribution in long-form generation.

\section{Synthetic Dataset Generation}
\label{dataset}

We constructed a diverse synthetic QA benchmark for the evaluation of our RAGentA framework using the DataMorgana platform~\cite{Filice2025DataMorgana}. The dataset comprises 500 QA pairs, each associated with multiple supporting evidence paragraphs. Detailed statistics of the dataset, including the distribution across various question and user categories, are presented in Table~\ref{tab:category_distribution}.

\begin{table}[h]
\centering
\caption{Distribution of Question and User Categories}
\label{tab:category_distribution}
\begin{tabular}{llr}
\toprule
                   Category Type &       Category Name &  Percent \\
\midrule
           Question: answer-type &          comparison &    49.8 \\
           Question: answer-type &        multi-aspect &    28.4 \\
           Question: answer-type &             factoid &    21.8 \\
\midrule
               Question: premise &     without premise &    67.4 \\
               Question: premise &        with premise &    32.6 \\
\midrule
           Question: formulation & concise and natural &    37.8 \\
           Question: formulation & verbose and natural &    35.0 \\
           Question: formulation &  short search query &    14.2 \\
           Question: formulation &   long search query &    13.0 \\
\midrule
            User: user-expertise &              expert &    50.8 \\
            User: user-expertise &              novice &    49.2 \\
\bottomrule
\end{tabular}
\end{table}
\section{The RAGentA Framework}
\label{methodology}
\begin{figure*}[htbp]
  \centering
  \includegraphics[width=\textwidth]{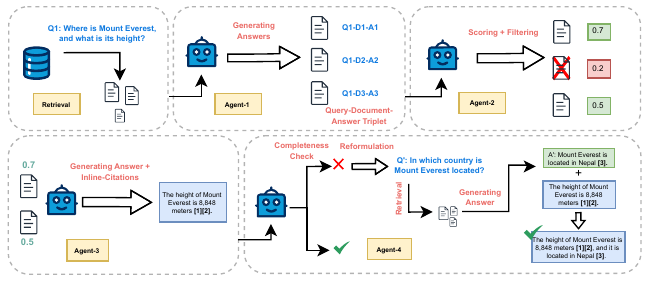}
  \caption{Architecture of the RAGentA framework: (1) A hybrid retriever selects top-20 documents. (2) Agent-1 generates an initial answer. (3) Agent-2 filters Query-Document-Answer triplets. (4) Agent-3 produces a final answer with in-line citations. (5) Agent-4 checks completeness, optionally reformulates the query, and merges both answers.}
  \Description{Visualization of the multi-agent RAG architecture of RAGentA.}
  \label{fig:architecture_RAGentA}
\end{figure*}

To address the SIGIR LiveRAG challenge, our primary objective is to ensure both the correctness and faithfulness of answers generated by RAG systems. To this end, we propose a multi-agent RAG architecture comprising four specialized agents, each powered by the Falcon-3-10B language model. The overall framework is illustrated in Figure~\ref{fig:architecture_RAGentA}. This approach is entirely training-free and broadly applicable to diverse QA tasks and domains without the need for additional fine-tuning or task-specific adaptation.

\subsection{Hybrid Retrieval System}
The first stage of our framework is a hybrid retrieval system that combines both dense and sparse models to ensure broad and contextually relevant document coverage. Sparse retrieval, based on BM25, is a widely adopted model that relies on exact term matching~\cite{Robertson1994BM25, Robertson1994Okapi}. Our semantic retrieval is powered by dense embeddings using the \textit{intfloat/e5-base-v2} model (E5) with Pinecone as the vector store to capture nuanced semantic relationships~\cite{Wang2024E5}. 

These two systems are fused using a tunable interpolation parameter $\alpha \in [0,1]$, producing a final document score as:

\begin{equation}
S_{\mathrm{hybrid}}(d) = \alpha \cdot S_{\mathrm{sparse}}(d) + (1 - \alpha) \cdot S_{\mathrm{dense}}(d)
\end{equation}

We set $\alpha = 0.35$ to give slightly higher weight to dense retrieval while maintaining meaningful contribution from sparse retrieval. This choice is motivated by the prior findings of~\citet{rayo2025hybrid}, where a hybrid retrieval with $\alpha = 0.35$ yielded the best performance. Their experiments on regulatory texts demonstrated that this balance improves overall retrieval performance.

\subsection{Multi-Agent Architecture}
Our multi-agent architecture builds up on the MAIN-RAG framework~\cite{Chang2024MAIN-RAG}, which employs a multi-agent system for document filtering and answer generation.  
In RAGentA, we preserve the core architecture of MAIN-RAG while introducing two key advancements: a substantial modification to Agent 3 and the addition of a novel Agent 4 dedicated to answer verification. Specifically, Agent 3 has been fundamentally redesigned to generate fine-grained in-line citations, thereby attributing each statement in the answer to its supporting evidence. In the following, we describe the roles and interactions of the four agents within the RAGentA framework.

\subsubsection{Agent-1: Predictor}  
Following document retrieval, Agent-1 generates query-specific answers for each retrieved document, forming Query-Document-Answer triplets. This agent employs an LLM to analyze each document in the context of the given query and generates an answer based on its content.

\subsubsection{Agent-2: Judge}  
This agent evaluates each document's relevance by determining its utility in addressing the query. For each triplet, Agent-2 addresses the question: “Is this document relevant and supportive for answering the question?” generating a binary “Yes” or “No” assessment. We implement adaptive document filtering by calculating a relevance score derived from the difference between logarithmic probabilities:
$\text{score} = \log p(\text{``Yes''}) - \log p(\text{``No''})$.
After scoring all documents, the system computes the mean score ($\tau_q$) and standard deviation ($\sigma$), establishing a dynamic threshold as $\text{adjusted\_}\tau_q = \tau_q - n \cdot \sigma$, where $n$ is a hyperparameter controlling filtering stringency. According to empirical analyses from the MAIN-RAG study, optimal performance is achieved at $n=0.5$. Utilizing this dynamic threshold, the system retains documents where $\text{score} \geq \text{adjusted\_}\tau_q$ and filters those where $\text{score} < \text{adjusted\_}\tau_q$.

\subsubsection{Agent-3: Final-Predictor} 
Following the filtering of irrelevant documents, Agent-3 synthesizes a comprehensive answer using the remaining, relevant documents. The principal modification to Agent-3, relative to MAIN-RAG, is the generation of explicit in-line citations. Specifically, Agent-3 is instructed to generate citations in the standardized $[X]$ format (where $X$ denotes the corresponding document identifier) using few-shot citation examples immediately after each factual assertion. As a result, each statement is treated as a discrete "claim", paired with its supporting citation(s), and extracted in a structured format. 
\subsubsection{Agent-4: Reviser}  
Agent-4 constitutes our most significant extension to the MAIN-RAG framework, conducting sophisticated multi-stage analysis to determine whether the query has been comprehensively answered. The reviser operates through a systematic pipeline that begins by decomposing complex queries (containing multiple sub-questions) into constituent components and mapping each generated claim to specific aspects of the original question. This process evaluates whether claims effectively address corresponding question components while filtering irrelevant ones.
Following the initial mapping, Agent-4 employs regular expression functions to parse structured output and assess answer completeness by categorizing each component as ``fully answered,'' ``partially answered,'' or ``not answered.'' When gaps are identified, Agent-4 initiates a targeted follow-up process: it generates questions to address these gaps and uses the hybrid retrieval system from initial phase, excluding previously retrieved documents to ensure complementary information. Agent-4 then generates answers for each follow-up question and integrates the new information into the original answer through answer synthesis. This iterative approach ensures complete coverage of all query components, resulting in a unified final answer that fully addresses the original question.
\section{Evaluation}
\label{evaluation}

We evaluate RAGentA with the 500 QA pairs of our synthetic dataset, which we described in Section~\ref{dataset}. The evaluation is a two-stage process, where we first assess retrieval performance and then evaluate the generated answers for correctness and faithfulness.

\subsection{Evaluation of Retrieval}
\label{evaluation_retrieval}

To assess the performance of our retrieval system, we employ two evaluation metrics: Recall@k and Mean Reciprocal Rank (MRR)@k, with k set to 20, as this corresponds to the number of documents provided to the RAGentA framework. Specifically, Recall@20 quantifies the proportion of relevant ground-truth evidence documents that are retrieved within the top 20 results, whereas MRR@20 measures the rank of the first relevant ground-truth document in the retrieved set. We compare the performance of our hybrid retrieval system to that of two baseline models, namely BM25 and E5, to evaluate its relative performance.

\begin{table}[h]
\centering
\caption{Evaluation of Retrieval Performance}
\label{tab:evaluation_retrieval}
\begin{tabular}{llr}
\toprule
                Retrieval System &               MRR@20 &        Recall@20 \\
\midrule
                            BM25 &               0.4205 &           0.5020 \\
                              E5 &               0.3476 &           0.4920 \\
                          Hybrid &      \textbf{0.4290} &  \textbf{0.5650} \\
\bottomrule
\end{tabular}
\end{table}

The evaluation results are presented in Table~\ref{tab:evaluation_retrieval}. Our hybrid retrieval system achieves an MRR@20 of 0.4290, outperforming BM25 (0.4205) by +2.0\% and E5 (0.3476) by +23.4\%. For Recall@20, the hybrid retrieval reaches 0.5650, exceeding BM25 (0.5020) by +12.5\% and E5 (0.4920) by +14.8\%. These findings demonstrate the strength of the hybrid retrieval approach, which fuses sparse and dense approaches in the RAGentA framework.

\subsection{Evaluation of Correctness and Faithfulness}

We aligned our evaluation of answer correctness and faithfulness closely with the autoevaluator used in the SIGIR LiveRAG challenge. Instead of the closed-source Claude-3.5 Sonnet model, we use \textit{Llama-3.3-70B-Instruct} as an LLM-as-a-judge, which receives the predicted answer, the ground-truth answer, and the cited passages. The LLM-as-a-judge assigns a \textit{correctness score} in the range \([-1, 2]\) and a \textit{faithfulness score} in the range \([-1, 1]\).

\subsubsection*{Correctness}
This metric consists of two components: \textit{coverage}, the portion of vital information in the ground-truth answer that is covered by the generated answer, inspired by the work of~\citet{Pradeep2025Correctness}, and \textit{relevance}, the portion of the generated answer that directly addresses the question, regardless of its factual correctness. Correctness is rated on a four-point scale:
\begin{itemize}
\item[\textbf{2}] The response correctly answers the user question and contains no irrelevant content.
\item[\textbf{1}] The response provides a useful answer to the user question, but may contain irrelevant content that does not harm the usefulness of the answer.
\item[\textbf{0}] No answer is provided in the response (e.g., ``I don’t know'').
\item[\textbf{-1}] The response does not answer the question whatsoever.
\end{itemize}

\subsubsection*{Faithfulness}
This metric assesses whether the answer is grounded in the retrieved documents on a three-point scale, following the methodology proposed by~\citet{Es2024Faithfulness}:
\begin{itemize}
\item[\textbf{1}] Full support: all answer parts are grounded.
\item[\textbf{0}] Partial support: not all answer parts are grounded.
\item[\textbf{-1}] No support: all answer parts are not grounded.
\end{itemize}

\begin{table}[h]
\centering
\caption{Evaluation of Correctness and Faithfulness}
\label{tab:evaluation_correctness_faithfullness}
\begin{tabular}{llr}
\toprule
                        Approach &          Correctness &     Faithfulness \\
\midrule
                    Standard RAG &               0.8256 &            0.6362 \\
          RAGentA \textit{(Ours)} &    \textbf{0.8346} &   \textbf{0.7044} \\
\bottomrule
\end{tabular}
\end{table}

Table~\ref{tab:evaluation_correctness_faithfullness} presents the average correctness and faithfulness scores computed over the full set of 500 questions. We evaluate the performance of RAGentA in comparison to a baseline RAG approach (Standard RAG). Both methods utilize the same set of initially retrieved documents to ensure a controlled comparison. Standard RAG does not incorporate agent-based reasoning and generates answers using a single prompt. The results show that RAGentA achieves a correctness score of 0.8348, slightly outperforming Standard RAG (0.8256) by +1.1\%. For faithfulness, RAGentA reaches 0.7044, exceeding Standard RAG (0.6362) by +10.7\%.

\section{Conclusion}
\label{conclusion}

In this work, we introduced RAGentA, a multi-agent RAG framework designed to enhance the trustworthiness of attributed QA. Our results demonstrate that a hybrid retrieval system combining BM25 and E5 significantly improves Recall@20 (+12.5\%) compared to the best single model. RAGentA outperforms the standard RAG baseline, particularly in faithfulness (+10.7\%), showing that providing in-line citations and conducting a second-stage retrieval for answer revision yield more relevant documents and well-grounded answers. While correctness sees only modest gains (+1.1\%), our analysis indicates that the second-stage retrieval currently offers limited added value and may benefit from further refinement, such as adding the document filtering by early-stage agents. Finally, we acknowledge that the four-agent design, while effective in boosting correctness, introduces substantial computational overhead. This trade-off highlights the need for future research to balance performance and efficiency in multi-agent RAG systems. 
The source code and experimental results presented in this study are publicly available in our Git repository at \url{https://github.com/faerber-lab/RAGentA}.

\begin{acks}
The authors acknowledge the financial support by the Federal Ministry of Research, Technology and Space of Germany and by Sächsische Staatsministerium für Wissenschaft, Kultur und Tourismus in the programme Center of Excellence for AI-research „Center for Scalable Data Analytics and Artificial Intelligence Dresden/Leipzig“, project identification number: ScaDS.AI.

The authors also acknowledge computing resources provided by the NHR Center at TU Dresden, supported by the Ministry of Research, Technology and Space and the participating state governments within the NHR framework.
\end{acks}

\bibliographystyle{ACM-Reference-Format}
\bibliography{literature}


\begin{thebibliography}{28}


\ifx \showCODEN    \undefined \def \showCODEN     #1{\unskip}     \fi
\ifx \showDOI      \undefined \def \showDOI       #1{#1}\fi
\ifx \showISBNx    \undefined \def \showISBNx     #1{\unskip}     \fi
\ifx \showISBNxiii \undefined \def \showISBNxiii  #1{\unskip}     \fi
\ifx \showISSN     \undefined \def \showISSN      #1{\unskip}     \fi
\ifx \showLCCN     \undefined \def \showLCCN      #1{\unskip}     \fi
\ifx \shownote     \undefined \def \shownote      #1{#1}          \fi
\ifx \showarticletitle \undefined \def \showarticletitle #1{#1}   \fi
\ifx \showURL      \undefined \def \showURL       {\relax}        \fi
\providecommand\bibfield[2]{#2}
\providecommand\bibinfo[2]{#2}
\providecommand\natexlab[1]{#1}
\providecommand\showeprint[2][]{arXiv:#2}

\bibitem[Berchansky et~al\mbox{.}(2024)]%
        {Berchansky2024CoTAR}
\bibfield{author}{\bibinfo{person}{Moshe Berchansky}, \bibinfo{person}{Daniel Fleischer}, \bibinfo{person}{Moshe Wasserblat}, {and} \bibinfo{person}{Peter Izsak}.} \bibinfo{year}{2024}\natexlab{}.
\newblock \showarticletitle{{C}o{TAR}: Chain-of-Thought Attribution Reasoning with Multi-level Granularity}. In \bibinfo{booktitle}{\emph{Findings of the Association for Computational Linguistics: EMNLP 2024}}. \bibinfo{publisher}{ACL}, \bibinfo{address}{Miami, Florida, USA}, \bibinfo{pages}{236--246}.
\newblock
\urldef\tempurl%
\url{https://doi.org/10.18653/v1/2024.findings-emnlp.13}
\showDOI{\tempurl}


\bibitem[Chang et~al\mbox{.}(2024)]%
        {Chang2024MAIN-RAG}
\bibfield{author}{\bibinfo{person}{Chia-Yuan Chang}, \bibinfo{person}{Zhimeng Jiang}, \bibinfo{person}{Vineeth Rakesh}, \bibinfo{person}{Menghai Pan}, \bibinfo{person}{Chin-Chia~Michael Yeh}, \bibinfo{person}{Guanchu Wang}, \bibinfo{person}{Mingzhi Hu}, \bibinfo{person}{Zhichao Xu}, \bibinfo{person}{Yan Zheng}, \bibinfo{person}{Mahashweta Das}, {and} \bibinfo{person}{Na Zou}.} \bibinfo{year}{2024}\natexlab{}.
\newblock \bibinfo{title}{MAIN-RAG: Multi-Agent Filtering Retrieval-Augmented Generation}.
\newblock
\newblock
\showeprint[arxiv]{2501.00332}~[cs.CL]


\bibitem[Es et~al\mbox{.}(2024)]%
        {Es2024Faithfulness}
\bibfield{author}{\bibinfo{person}{Shahul Es}, \bibinfo{person}{Jithin James}, \bibinfo{person}{Luis Espinosa~Anke}, {and} \bibinfo{person}{Steven Schockaert}.} \bibinfo{year}{2024}\natexlab{}.
\newblock \showarticletitle{{RAGA}s: Automated Evaluation of Retrieval Augmented Generation}. In \bibinfo{booktitle}{\emph{Proceedings of the 18th Conference of the European Chapter of the Association for Computational Linguistics: System Demonstrations}}. \bibinfo{publisher}{ACL}, \bibinfo{address}{St. Julians, Malta}, \bibinfo{pages}{150--158}.
\newblock
\urldef\tempurl%
\url{https://aclanthology.org/2024.eacl-demo.16/}
\showURL{%
\tempurl}


\bibitem[Fierro et~al\mbox{.}(2024)]%
        {Fierro2024LearningToPlan}
\bibfield{author}{\bibinfo{person}{Constanza Fierro}, \bibinfo{person}{Reinald~Kim Amplayo}, \bibinfo{person}{Fantine Huot}, \bibinfo{person}{Nicola De~Cao}, \bibinfo{person}{Joshua Maynez}, \bibinfo{person}{Shashi Narayan}, {and} \bibinfo{person}{Mirella Lapata}.} \bibinfo{year}{2024}\natexlab{}.
\newblock \showarticletitle{Learning to Plan and Generate Text with Citations}. In \bibinfo{booktitle}{\emph{Proceedings of the 62nd Annual Meeting of the Association for Computational Linguistics (Volume 1: Long Papers)}}. \bibinfo{publisher}{ACL}, \bibinfo{address}{Bangkok, Thailand}, \bibinfo{pages}{11397--11417}.
\newblock
\urldef\tempurl%
\url{https://doi.org/10.18653/v1/2024.acl-long.615}
\showDOI{\tempurl}


\bibitem[Filice et~al\mbox{.}(2025)]%
        {Filice2025DataMorgana}
\bibfield{author}{\bibinfo{person}{Simone Filice}, \bibinfo{person}{Guy Horowitz}, \bibinfo{person}{David Carmel}, \bibinfo{person}{Zohar Karnin}, \bibinfo{person}{Liane Lewin-Eytan}, {and} \bibinfo{person}{Yoelle Maarek}.} \bibinfo{year}{2025}\natexlab{}.
\newblock \bibinfo{title}{Generating Diverse Q\&A Benchmarks for RAG Evaluation with DataMorgana}.
\newblock
\newblock
\showeprint[arxiv]{2501.12789}~[cs.CL]


\bibitem[Gao et~al\mbox{.}(2023a)]%
        {Gao2023RARR}
\bibfield{author}{\bibinfo{person}{Luyu Gao}, \bibinfo{person}{Zhuyun Dai}, \bibinfo{person}{Panupong Pasupat}, \bibinfo{person}{Anthony Chen}, \bibinfo{person}{Arun~Tejasvi Chaganty}, \bibinfo{person}{Yicheng Fan}, \bibinfo{person}{Vincent Zhao}, \bibinfo{person}{Ni Lao}, \bibinfo{person}{Hongrae Lee}, \bibinfo{person}{Da-Cheng Juan}, {and} \bibinfo{person}{Kelvin Guu}.} \bibinfo{year}{2023}\natexlab{a}.
\newblock \showarticletitle{{RARR}: Researching and Revising What Language Models Say, Using Language Models}. In \bibinfo{booktitle}{\emph{Proceedings of the 61st Annual Meeting of the Association for Computational Linguistics (Volume 1: Long Papers)}}. \bibinfo{publisher}{ACL}, \bibinfo{address}{Toronto, Canada}, \bibinfo{pages}{16477--16508}.
\newblock
\urldef\tempurl%
\url{https://doi.org/10.18653/v1/2023.acl-long.910}
\showDOI{\tempurl}


\bibitem[Gao et~al\mbox{.}(2023b)]%
        {Gao2023ALCE}
\bibfield{author}{\bibinfo{person}{Tianyu Gao}, \bibinfo{person}{Howard Yen}, \bibinfo{person}{Jiatong Yu}, {and} \bibinfo{person}{Danqi Chen}.} \bibinfo{year}{2023}\natexlab{b}.
\newblock \showarticletitle{Enabling Large Language Models to Generate Text with Citations}. In \bibinfo{booktitle}{\emph{Proceedings of the 2023 Conference on Empirical Methods in Natural Language Processing}}. \bibinfo{publisher}{ACL}, \bibinfo{address}{Singapore}, \bibinfo{pages}{6465--6488}.
\newblock
\urldef\tempurl%
\url{https://doi.org/10.18653/v1/2023.emnlp-main.398}
\showDOI{\tempurl}


\bibitem[Huang et~al\mbox{.}(2024)]%
        {Huang2024CitationReward}
\bibfield{author}{\bibinfo{person}{Chengyu Huang}, \bibinfo{person}{Zeqiu Wu}, \bibinfo{person}{Yushi Hu}, {and} \bibinfo{person}{Wenya Wang}.} \bibinfo{year}{2024}\natexlab{}.
\newblock \showarticletitle{Training Language Models to Generate Text with Citations via Fine-grained Rewards}. In \bibinfo{booktitle}{\emph{Proceedings of the 62nd Annual Meeting of the Association for Computational Linguistics (Volume 1: Long Papers)}}. \bibinfo{publisher}{ACL}, \bibinfo{address}{Bangkok, Thailand}, \bibinfo{pages}{2926--2949}.
\newblock
\urldef\tempurl%
\url{https://doi.org/10.18653/v1/2024.acl-long.161}
\showDOI{\tempurl}


\bibitem[Huang and Chang(2024)]%
        {Huang2024Citation}
\bibfield{author}{\bibinfo{person}{Jie Huang} {and} \bibinfo{person}{Kevin Chang}.} \bibinfo{year}{2024}\natexlab{}.
\newblock \showarticletitle{Citation: A Key to Building Responsible and Accountable Large Language Models}. In \bibinfo{booktitle}{\emph{Findings of the Association for Computational Linguistics: NAACL 2024}}. \bibinfo{publisher}{ACL}, \bibinfo{address}{Mexico City, Mexico}, \bibinfo{pages}{464--473}.
\newblock
\urldef\tempurl%
\url{https://doi.org/10.18653/v1/2024.findings-naacl.31}
\showDOI{\tempurl}


\bibitem[Huang et~al\mbox{.}(2025)]%
        {Huang2025Hallucination}
\bibfield{author}{\bibinfo{person}{Lei Huang}, \bibinfo{person}{Weijiang Yu}, \bibinfo{person}{Weitao Ma}, \bibinfo{person}{Weihong Zhong}, \bibinfo{person}{Zhangyin Feng}, \bibinfo{person}{Haotian Wang}, \bibinfo{person}{Qianglong Chen}, \bibinfo{person}{Weihua Peng}, \bibinfo{person}{Xiaocheng Feng}, \bibinfo{person}{Bing Qin}, {and} \bibinfo{person}{Ting Liu}.} \bibinfo{year}{2025}\natexlab{}.
\newblock \showarticletitle{A Survey on Hallucination in Large Language Models: Principles, Taxonomy, Challenges, and Open Questions}.
\newblock \bibinfo{journal}{\emph{ACM Trans. Inf. Syst.}} \bibinfo{volume}{43}, \bibinfo{number}{2}, Article \bibinfo{articleno}{42} (\bibinfo{year}{2025}), \bibinfo{numpages}{55}~pages.
\newblock
\urldef\tempurl%
\url{https://doi.org/10.1145/3703155}
\showDOI{\tempurl}


\bibitem[Jang and Li(2024)]%
        {Jang2024AU-RAG}
\bibfield{author}{\bibinfo{person}{Jisoo Jang} {and} \bibinfo{person}{Wen-Syan Li}.} \bibinfo{year}{2024}\natexlab{}.
\newblock \showarticletitle{AU-RAG: Agent-based Universal Retrieval Augmented Generation}. In \bibinfo{booktitle}{\emph{Proceedings of the 2024 Annual International ACM SIGIR Conference on Research and Development in Information Retrieval in the Asia Pacific Region}} (Tokyo, Japan) \emph{(\bibinfo{series}{SIGIR-AP 2024})}. \bibinfo{publisher}{ACM}, \bibinfo{address}{New York, NY, USA}, \bibinfo{pages}{2–11}.
\newblock
\urldef\tempurl%
\url{https://doi.org/10.1145/3673791.3698416}
\showDOI{\tempurl}


\bibitem[Lewis et~al\mbox{.}(2020)]%
        {Lewis2020RAG}
\bibfield{author}{\bibinfo{person}{Patrick Lewis}, \bibinfo{person}{Ethan Perez}, \bibinfo{person}{Aleksandra Piktus}, \bibinfo{person}{Fabio Petroni}, \bibinfo{person}{Vladimir Karpukhin}, \bibinfo{person}{Naman Goyal}, \bibinfo{person}{Heinrich K\"{u}ttler}, \bibinfo{person}{Mike Lewis}, \bibinfo{person}{Wen-tau Yih}, \bibinfo{person}{Tim Rockt\"{a}schel}, \bibinfo{person}{Sebastian Riedel}, {and} \bibinfo{person}{Douwe Kiela}.} \bibinfo{year}{2020}\natexlab{}.
\newblock \showarticletitle{Retrieval-Augmented Generation for Knowledge-Intensive NLP Tasks}. In \bibinfo{booktitle}{\emph{Advances in Neural Information Processing Systems}}, Vol.~\bibinfo{volume}{33}. \bibinfo{publisher}{Curran Associates, Inc.}, \bibinfo{address}{Red Hook, NY, USA}, \bibinfo{pages}{9459--9474}.
\newblock
\urldef\tempurl%
\url{https://dl.acm.org/doi/abs/10.5555/3495724.3496517}
\showURL{%
\tempurl}


\bibitem[Mosquera et~al\mbox{.}(2025)]%
        {rayo2025hybrid}
\bibfield{author}{\bibinfo{person}{Jhon Stewar~Rayo Mosquera}, \bibinfo{person}{Carlos Raúl De La~Rosa Peredo}, {and} \bibinfo{person}{Mario~Garrido Córdoba}.} \bibinfo{year}{2025}\natexlab{}.
\newblock \showarticletitle{A Hybrid Approach to Information Retrieval and Answer Generation for Regulatory Texts}. In \bibinfo{booktitle}{\emph{Proceedings of the 1st Regulatory NLP Workshop (RegNLP 2025)}}. \bibinfo{publisher}{ACL}, \bibinfo{address}{Abu Dhabi, UAE}, \bibinfo{pages}{31--35}.
\newblock
\urldef\tempurl%
\url{https://aclanthology.org/2025.regnlp-1.5/}
\showURL{%
\tempurl}


\bibitem[Patel et~al\mbox{.}(2024)]%
        {Patel2024MultiSource}
\bibfield{author}{\bibinfo{person}{Nilay Patel}, \bibinfo{person}{Shivashankar Subramanian}, \bibinfo{person}{Siddhant Garg}, \bibinfo{person}{Pratyay Banerjee}, {and} \bibinfo{person}{Amita Misra}.} \bibinfo{year}{2024}\natexlab{}.
\newblock \showarticletitle{Towards Improved Multi-Source Attribution for Long-Form Answer Generation}. In \bibinfo{booktitle}{\emph{Proceedings of the 2024 Conference of the North American Chapter of the Association for Computational Linguistics: Human Language Technologies (Volume 1: Long Papers)}}. \bibinfo{publisher}{ACL}, \bibinfo{address}{Mexico City, Mexico}, \bibinfo{pages}{3906--3919}.
\newblock
\urldef\tempurl%
\url{https://doi.org/10.18653/v1/2024.naacl-long.216}
\showDOI{\tempurl}


\bibitem[Pradeep et~al\mbox{.}(2025)]%
        {Pradeep2025Correctness}
\bibfield{author}{\bibinfo{person}{Ronak Pradeep}, \bibinfo{person}{Nandan Thakur}, \bibinfo{person}{Shivani Upadhyay}, \bibinfo{person}{Daniel Campos}, \bibinfo{person}{Nick Craswell}, {and} \bibinfo{person}{Jimmy Lin}.} \bibinfo{year}{2025}\natexlab{}.
\newblock \bibinfo{title}{The Great Nugget Recall: Automating Fact Extraction and RAG Evaluation with Large Language Models}.
\newblock
\newblock
\showeprint[arxiv]{2504.15068}~[cs.IR]


\bibitem[Qi et~al\mbox{.}(2024)]%
        {Qi2024ModelInternals}
\bibfield{author}{\bibinfo{person}{Jirui Qi}, \bibinfo{person}{Gabriele Sarti}, \bibinfo{person}{Raquel Fern{\'a}ndez}, {and} \bibinfo{person}{Arianna Bisazza}.} \bibinfo{year}{2024}\natexlab{}.
\newblock \showarticletitle{Model Internals-based Answer Attribution for Trustworthy Retrieval-Augmented Generation}. In \bibinfo{booktitle}{\emph{Proceedings of the 2024 Conference on Empirical Methods in Natural Language Processing}}. \bibinfo{publisher}{ACL}, \bibinfo{address}{Miami, Florida, USA}, \bibinfo{pages}{6037--6053}.
\newblock
\urldef\tempurl%
\url{https://doi.org/10.18653/v1/2024.emnlp-main.347}
\showDOI{\tempurl}


\bibitem[Ramu et~al\mbox{.}(2024)]%
        {Ramu2024PostHoc}
\bibfield{author}{\bibinfo{person}{Pritika Ramu}, \bibinfo{person}{Koustava Goswami}, \bibinfo{person}{Apoorv Saxena}, {and} \bibinfo{person}{Balaji~Vasan Srinivasan}.} \bibinfo{year}{2024}\natexlab{}.
\newblock \showarticletitle{Enhancing Post-Hoc Attributions in Long Document Comprehension via Coarse Grained Answer Decomposition}. In \bibinfo{booktitle}{\emph{Proceedings of the 2024 Conference on Empirical Methods in Natural Language Processing}}. \bibinfo{publisher}{ACL}, \bibinfo{address}{Miami, Florida, USA}, \bibinfo{pages}{17790--17806}.
\newblock
\urldef\tempurl%
\url{https://doi.org/10.18653/v1/2024.emnlp-main.985}
\showDOI{\tempurl}


\bibitem[Robertson and Walker(1994)]%
        {Robertson1994BM25}
\bibfield{author}{\bibinfo{person}{S.~E. Robertson} {and} \bibinfo{person}{S. Walker}.} \bibinfo{year}{1994}\natexlab{}.
\newblock \showarticletitle{Some simple effective approximations to the 2-Poisson model for probabilistic weighted retrieval}. In \bibinfo{booktitle}{\emph{ACM SIGIR'94}} (Dublin, Ireland). \bibinfo{publisher}{Springer-Verlag}, \bibinfo{address}{Berlin, Heidelberg}, \bibinfo{pages}{232–241}.
\newblock
\showISBNx{038719889X}
\urldef\tempurl%
\url{https://doi.org/10.1007/978-1-4471-2099-5_24}
\showDOI{\tempurl}


\bibitem[Robertson et~al\mbox{.}(1994)]%
        {Robertson1994Okapi}
\bibfield{author}{\bibinfo{person}{Stephen~E. Robertson}, \bibinfo{person}{Steve Walker}, \bibinfo{person}{Susan Jones}, \bibinfo{person}{Micheline Hancock{-}Beaulieu}, {and} \bibinfo{person}{Mike Gatford}.} \bibinfo{year}{1994}\natexlab{}.
\newblock \showarticletitle{Okapi at {TREC-3}}. In \bibinfo{booktitle}{\emph{TREC'94}}, Vol.~\bibinfo{volume}{500-225}. \bibinfo{publisher}{NIST}, \bibinfo{address}{Gaithersburg, USA}, \bibinfo{pages}{109--126}.
\newblock
\urldef\tempurl%
\url{http://trec.nist.gov/pubs/trec3/papers/city.ps.gz}
\showURL{%
\tempurl}


\bibitem[Saeid and Kopinski(2024)]%
        {Saeid2024AgentFusion}
\bibfield{author}{\bibinfo{person}{Yasser Saeid} {and} \bibinfo{person}{Thomas Kopinski}.} \bibinfo{year}{2024}\natexlab{}.
\newblock \showarticletitle{AgentFusion: A Multi-Agent Approach to Accurate Text Generation}. In \bibinfo{booktitle}{\emph{2024 International Conference on Electrical and Computer Engineering Researches (ICECER)}}. \bibinfo{publisher}{IEEE}, \bibinfo{address}{Gaborone, Botswana}, \bibinfo{pages}{1--8}.
\newblock
\urldef\tempurl%
\url{https://doi.org/10.1109/ICECER62944.2024.10920460}
\showDOI{\tempurl}


\bibitem[Singhal et~al\mbox{.}(2025)]%
        {Singhal2025QA}
\bibfield{author}{\bibinfo{person}{Karan Singhal}, \bibinfo{person}{Tao Tu}, \bibinfo{person}{Juraj Gottweis}, \bibinfo{person}{Rory Sayres}, \bibinfo{person}{Ellery Wulczyn}, \bibinfo{person}{Mohamed Amin}, \bibinfo{person}{Le Hou}, \bibinfo{person}{Kevin Clark}, \bibinfo{person}{Stephen~R. Pfohl}, \bibinfo{person}{Heather Cole-Lewis}, \bibinfo{person}{Darlene Neal}, \bibinfo{person}{Qazi~Mamunur Rashid}, \bibinfo{person}{Mike Schaekermann}, \bibinfo{person}{Amy Wang}, \bibinfo{person}{Dev Dash}, \bibinfo{person}{Jonathan~H. Chen}, \bibinfo{person}{Nigam~H. Shah}, \bibinfo{person}{Sami Lachgar}, \bibinfo{person}{Philip~Andrew Mansfield}, \bibinfo{person}{Sushant Prakash}, \bibinfo{person}{Bradley Green}, \bibinfo{person}{Ewa Dominowska}, \bibinfo{person}{Blaise Ag{\"u}era~y Arcas}, \bibinfo{person}{Nenad Toma{\v{s}}ev}, \bibinfo{person}{Yun Liu}, \bibinfo{person}{Renee Wong}, \bibinfo{person}{Christopher Semturs}, \bibinfo{person}{S.~Sara Mahdavi}, \bibinfo{person}{Joelle~K. Barral}, \bibinfo{person}{Dale~R.
  Webster}, \bibinfo{person}{Greg~S. Corrado}, \bibinfo{person}{Yossi Matias}, \bibinfo{person}{Shekoofeh Azizi}, \bibinfo{person}{Alan Karthikesalingam}, {and} \bibinfo{person}{Vivek Natarajan}.} \bibinfo{year}{2025}\natexlab{}.
\newblock \showarticletitle{Toward expert-level medical question answering with large language models}.
\newblock \bibinfo{journal}{\emph{Nature Medicine}} \bibinfo{volume}{31}, \bibinfo{number}{3} (\bibinfo{date}{01 Mar} \bibinfo{year}{2025}), \bibinfo{pages}{943--950}.
\newblock
\showISSN{1546-170X}
\urldef\tempurl%
\url{https://doi.org/10.1038/s41591-024-03423-7}
\showDOI{\tempurl}


\bibitem[Taylor et~al\mbox{.}(2022)]%
        {Taylor2022Galactica}
\bibfield{author}{\bibinfo{person}{Ross Taylor}, \bibinfo{person}{Marcin Kardas}, \bibinfo{person}{Guillem Cucurull}, \bibinfo{person}{Thomas Scialom}, \bibinfo{person}{Anthony Hartshorn}, \bibinfo{person}{Elvis Saravia}, \bibinfo{person}{Andrew Poulton}, \bibinfo{person}{Viktor Kerkez}, {and} \bibinfo{person}{Robert Stojnic}.} \bibinfo{year}{2022}\natexlab{}.
\newblock \bibinfo{title}{Galactica: A Large Language Model for Science}.
\newblock
\newblock
\showeprint[arxiv]{2211.09085}~[cs.CL]


\bibitem[Wang et~al\mbox{.}(2024)]%
        {Wang2024E5}
\bibfield{author}{\bibinfo{person}{Liang Wang}, \bibinfo{person}{Nan Yang}, \bibinfo{person}{Xiaolong Huang}, \bibinfo{person}{Binxing Jiao}, \bibinfo{person}{Linjun Yang}, \bibinfo{person}{Daxin Jiang}, \bibinfo{person}{Rangan Majumder}, {and} \bibinfo{person}{Furu Wei}.} \bibinfo{year}{2024}\natexlab{}.
\newblock \bibinfo{title}{Text Embeddings by Weakly-Supervised Contrastive Pre-training}.
\newblock
\newblock
\showeprint[arxiv]{2212.03533}~[cs.CL]


\bibitem[Wei et~al\mbox{.}(2022)]%
        {Wei2022CoT}
\bibfield{author}{\bibinfo{person}{Jason Wei}, \bibinfo{person}{Xuezhi Wang}, \bibinfo{person}{Dale Schuurmans}, \bibinfo{person}{Maarten Bosma}, \bibinfo{person}{Brian Ichter}, \bibinfo{person}{Fei Xia}, \bibinfo{person}{Ed~H. Chi}, \bibinfo{person}{Quoc~V. Le}, {and} \bibinfo{person}{Denny Zhou}.} \bibinfo{year}{2022}\natexlab{}.
\newblock \showarticletitle{Chain-of-thought prompting elicits reasoning in large language models}. In \bibinfo{booktitle}{\emph{Proceedings of the 36th International Conference on Neural Information Processing Systems}} (New Orleans, LA, USA) \emph{(\bibinfo{series}{NIPS '22})}. \bibinfo{publisher}{Curran Associates Inc.}, \bibinfo{address}{Red Hook, NY, USA}, Article \bibinfo{articleno}{1800}, \bibinfo{numpages}{14}~pages.
\newblock
\showISBNx{9781713871088}
\urldef\tempurl%
\url{https://dl.acm.org/doi/10.5555/3600270.3602070}
\showURL{%
\tempurl}


\bibitem[Yang et~al\mbox{.}(2024)]%
        {Yang2024IM-RAG}
\bibfield{author}{\bibinfo{person}{Diji Yang}, \bibinfo{person}{Jinmeng Rao}, \bibinfo{person}{Kezhen Chen}, \bibinfo{person}{Xiaoyuan Guo}, \bibinfo{person}{Yawen Zhang}, \bibinfo{person}{Jie Yang}, {and} \bibinfo{person}{Yi Zhang}.} \bibinfo{year}{2024}\natexlab{}.
\newblock \showarticletitle{IM-RAG: Multi-Round Retrieval-Augmented Generation Through Learning Inner Monologues}. In \bibinfo{booktitle}{\emph{Proceedings of the 47th International ACM SIGIR Conference on Research and Development in Information Retrieval}} (Washington DC, USA) \emph{(\bibinfo{series}{SIGIR '24})}. \bibinfo{publisher}{ACM}, \bibinfo{address}{New York, NY, USA}, \bibinfo{pages}{730–740}.
\newblock
\showISBNx{9798400704314}
\urldef\tempurl%
\url{https://doi.org/10.1145/3626772.3657760}
\showDOI{\tempurl}


\bibitem[Zhang et~al\mbox{.}(2024)]%
        {Zhang2024Summarization}
\bibfield{author}{\bibinfo{person}{Tianyi Zhang}, \bibinfo{person}{Faisal Ladhak}, \bibinfo{person}{Esin Durmus}, \bibinfo{person}{Percy Liang}, \bibinfo{person}{Kathleen McKeown}, {and} \bibinfo{person}{Tatsunori~B. Hashimoto}.} \bibinfo{year}{2024}\natexlab{}.
\newblock \showarticletitle{Benchmarking Large Language Models for News Summarization}.
\newblock \bibinfo{journal}{\emph{Transactions of the Association for Computational Linguistics}}  \bibinfo{volume}{12} (\bibinfo{year}{2024}), \bibinfo{pages}{39--57}.
\newblock
\urldef\tempurl%
\url{https://doi.org/10.1162/tacl_a_00632}
\showDOI{\tempurl}


\bibitem[Zhao et~al\mbox{.}(2024)]%
        {Zhao2024Longagent}
\bibfield{author}{\bibinfo{person}{Jun Zhao}, \bibinfo{person}{Can Zu}, \bibinfo{person}{Xu Hao}, \bibinfo{person}{Yi Lu}, \bibinfo{person}{Wei He}, \bibinfo{person}{Yiwen Ding}, \bibinfo{person}{Tao Gui}, \bibinfo{person}{Qi Zhang}, {and} \bibinfo{person}{Xuanjing Huang}.} \bibinfo{year}{2024}\natexlab{}.
\newblock \showarticletitle{{LONGAGENT}: Achieving Question Answering for 128k-Token-Long Documents through Multi-Agent Collaboration}. In \bibinfo{booktitle}{\emph{Proceedings of the 2024 Conference on Empirical Methods in Natural Language Processing}}. \bibinfo{publisher}{ACL}, \bibinfo{address}{Miami, Florida, USA}, \bibinfo{pages}{16310--16324}.
\newblock
\urldef\tempurl%
\url{https://doi.org/10.18653/v1/2024.emnlp-main.912}
\showDOI{\tempurl}


\bibitem[Zhu et~al\mbox{.}(2024)]%
        {Zhu2024ATM}
\bibfield{author}{\bibinfo{person}{Junda Zhu}, \bibinfo{person}{Lingyong Yan}, \bibinfo{person}{Haibo Shi}, \bibinfo{person}{Dawei Yin}, {and} \bibinfo{person}{Lei Sha}.} \bibinfo{year}{2024}\natexlab{}.
\newblock \showarticletitle{{ATM}: Adversarial Tuning Multi-agent System Makes a Robust Retrieval-Augmented Generator}. In \bibinfo{booktitle}{\emph{Proceedings of the 2024 Conference on Empirical Methods in Natural Language Processing}}. \bibinfo{publisher}{ACL}, \bibinfo{address}{Miami, Florida, USA}, \bibinfo{pages}{10902--10919}.
\newblock
\urldef\tempurl%
\url{https://doi.org/10.18653/v1/2024.emnlp-main.610}
\showDOI{\tempurl}


\end{thebibliography}

\end{document}